# Near-Infrared Emitting Single Squaraine Dye Aggregate with large Stokes shift

G. M. Paternò[1], L. Moretti[2], A. Barker[1], N. Barbero[3], S. Galliano[3], C. Barolo[3,4], G. Lanzani[1,2] and F. Scotognella[1,2]

[1] Center for Nano Science and Technology@PoliMi, Istituto Italiano di Tecnologia, Via Giovanni Pascoli, 70/3, 20133, Milan, Italy

[2] Dipartimento di Fisica, Politecnico di Milano, Piazza Leonardo da Vinci 32, 20133 Milano, Italy

[3] Dipartimento di Chimica and NIS Interdepartmental and INSTM Reference Centre, Università degli Studi di Torino, Via Pietro Giuria 7, 10125 Torino, Italy

[4] ICxT Interdepartmental Centre, Università di Torino, Lungo Dora Siena 100, 10100 Torino, Italy

## Abstract

The study of supramolecular interactions and aggregation behaviour of functional materials is of great importance to tune and extend their spectral sensitivity and, hence, improve the optoelectronic response of related devices. In this study, we resolve spatially and spectrally the absorption and emission features of a squaraine aggregate by means of confocal microscopy and absorption/photoluminescence spectroscopy. We observe that the aggregate affords both a broad absorption spectrum (centred at 670 nm), likely originated by a dyes configuration with allowed J- and H- arrangements, and a strong and relatively narrow emission in the near-infrared (NIR) part of the spectrum (centred at 780 nm), with a remarkable Stokes shift of 110 nm that is among the largest exhibited by squaraine dyes. These peculiarities would be beneficial for extending the spectral sensitivity of bot photovoltaic and light-emitting diodes, and extremely appealing for possible applications of these aggregates as NIR fluorescent probes in biomedical applications.

# Introduction

The optical properties and aggregation behaviour of squaraine dyes (SQ) have attracted a wide research interest in recent years, owing to their effective application in dye-sensitised,[1-3] organic[4,5] and perovskite solar cells,[6] as well as in organic light-emitting diodes (OLEDs)[7] and biomedical applications.[8,9] In particular, given their relatively strong light absorption ($\epsilon > 300000$ mol$^{-1}$ cm$^{-1}$) and emission features in the near-infrared (NIR) part of the spectrum, SQs represent a class of the materials well suited for harvesting/emitting photons in the NIR region.

In this context, supramolecular interaction[10] is another important factor that plays a prominent role in tuning the spectral response of organic functional materials and, hence, widening up their possible range of applications. For these reasons, the study of self-organisation, crystallisation and aggregation behaviour of a variety of functional materials and chromophores have become an important topic in recent years.[11-17] More specifically, SQ dyes show a strong tendency to form both J-type and H-type aggregates in solution[16], with the former featuring a significant bathochromic shift and enhancement of absorption and the latter a hypsochromic shift and decrease of absorption with the respect of the monomeric band. On the other hand, in solid state[18-21] the presence of polymorphic species results in the observation of a complex aggregation pattern that includes the coexistence of J-, H- and also "oblique aggregates" that lead to a splitting of the absorption band into two components around the monomer absorption, the so-called Davydov splitting.[22] Therefore, it is somehow difficult to investigate the various aggregation contributions at the micro/nanoscale and map out the molecular arrangement in solid state. The study of the intrinsic optical properties of aggregates in solid films and polycrystalline materials would be, in fact, of great practical interest for the effective utilisation of such dyes in optoelectronic devices and biosensing/imaging. This can be more relevant, for instance, in OPVs for which blending with electron acceptor materials increases the degree of morphological complexity of the active layer, and usually leads to a more convoluted aggregation pattern of SQs[18,23].

Herein, we report a study of a single micron-sized aggregate of a squaraine derivative, namely VG1 – C8, exhibiting large Stokes shift (110 nm) and a considerable spectral distance between excitation (570 nm) and emission (780 nm) wavelengths that are among the highest reported for SQs[24]. By probing the different domains of the aggregate and of the surrounding film by means of confocal microscopy we observed that the aggregate, of likely polycrystalline origin, shows a spectrally extended absorption both on the lower and higher energy sides of the monomer peak, possibly due to the coexistence of both J- and H- aggregates or Davydov splitting. Interestingly, we found that the photoluminescence from the aggregate core is substantially shifted towards the NIR region (780 nm), suggesting that an appreciable fraction of aggregates affords a highly emissive J- configuration. These properties would permit to extend the spectral sensitivity of optoelectronic devices (i.e. OPVs and OLEDs) and would be ideal for possible applications in biosensing and bioimaging.[25]

## Experimental

### Synthesis

5-carboxy-2,3,3-trimethyl-1-octyl-3H-indol-1-ium iodide (VG1-C8) was synthesised as previously reported in the literature[1, 26].

### Aggregates formation

The aggregates, with sizes ranging from 20 μm to 80 μm, formed spontaneously upon increase of the concentration (10 mg/mL, 1.4 mM) in ethanol. To investigate the properties of those aggregates, we drop-cast the obtained dispersion, which contained both the aggregates and the saturated solution. This allowed us to study both the aggregate and the surrounding film.

### Confocal absorption and photoluminescence

The set-up consisted of a homemade microscope in transmission configuration, equipped with an 100× objective with NA of 0.7. The probe and excitation light came from a NKT Photonics SuperK Select laser, continuously tunable from 475 to 1050 nm. The transmitted light was detected with an optical fibre connected to a silicon-based photodiode. Transmittance was calculated as $T = I/I_0$ by recording $I_0$ in a featureless region, and I in the region of interest. Light transmission and photoluminescence were spatially resolved by raster-scanning the sample with respect to the laser beam through a piezo stage (P-517 Physik Instrumente) in x-y plane, with the z excursion used to optimise the focus on the sample. Transmission maps were taken at 500 nm, 690 nm and 880 nm, while localised transmission spectra were collected by scanning the laser emission wavelength. Photoluminescence (PL) measurements were performed by exciting the sample at 570 nm. The collected PL wavelength was selected by inserting low-pass (LP) or band pass (BP) filters between the sample and the optical fibre coupler, enabling the collection of PL maps that spatially resolve different aggregate emissions. Localised PL spectra were recorded by coupling the collection fibre to an Ocean Optics Maya2000 Pro spectrometer, using a 600nm LP filter to remove the residual excitation light.

## Results and discussion

Figure 1 summarises the main molecular and optical properties of VG1-C8, alongside with a visual description of the aggregates formation. The absorption spectrum of VG1-C8 in dilute solution (0.1 mg/mL, 0.14 mM) shows a strong and narrow band at 645 nm that can be attributed to the $\pi \rightarrow \pi^*$ transition, and its vibronic shoulder at 600 nm.[3] The small Stokes shift of VG1–C8 (9 nm) is typical for symmetric squaraine derivatives, and it is indicative of comparable molecular configuration between the ground and excited electronic states of the molecule.[27, 28] On the other hand, we noticed the spontaneous formation of shiny aggregates upon increase of the concentration to 10 mg/mL (14.03 mM) in ethanol (fig. 1c), with sizes ranging from 20 μm to 80 μm as measured by optical microscopy (fig. 1d) and thickness ranging from ≈ 200 nm to ≈ 600 nm (by profilometry). Although the geometrical regularity and brilliance suggested a possible single crystalline nature of the aggregates, their single crystal X-rays characterisation did not give any appreciable scattering diffraction pattern, thus suggesting that they are likely of polycrystalline origin (for powder diffraction pattern of VG1 –C8 see ref [29]). We therefore drop-cast the VG1–C8 aggregates dispersed in the ethanol solution on a glass substrate, in order to proceed to the spectroscopic investigation of the squaraine aggregate and the surrounding film via confocal microscopy and absorption/photoluminescence spectroscopy.

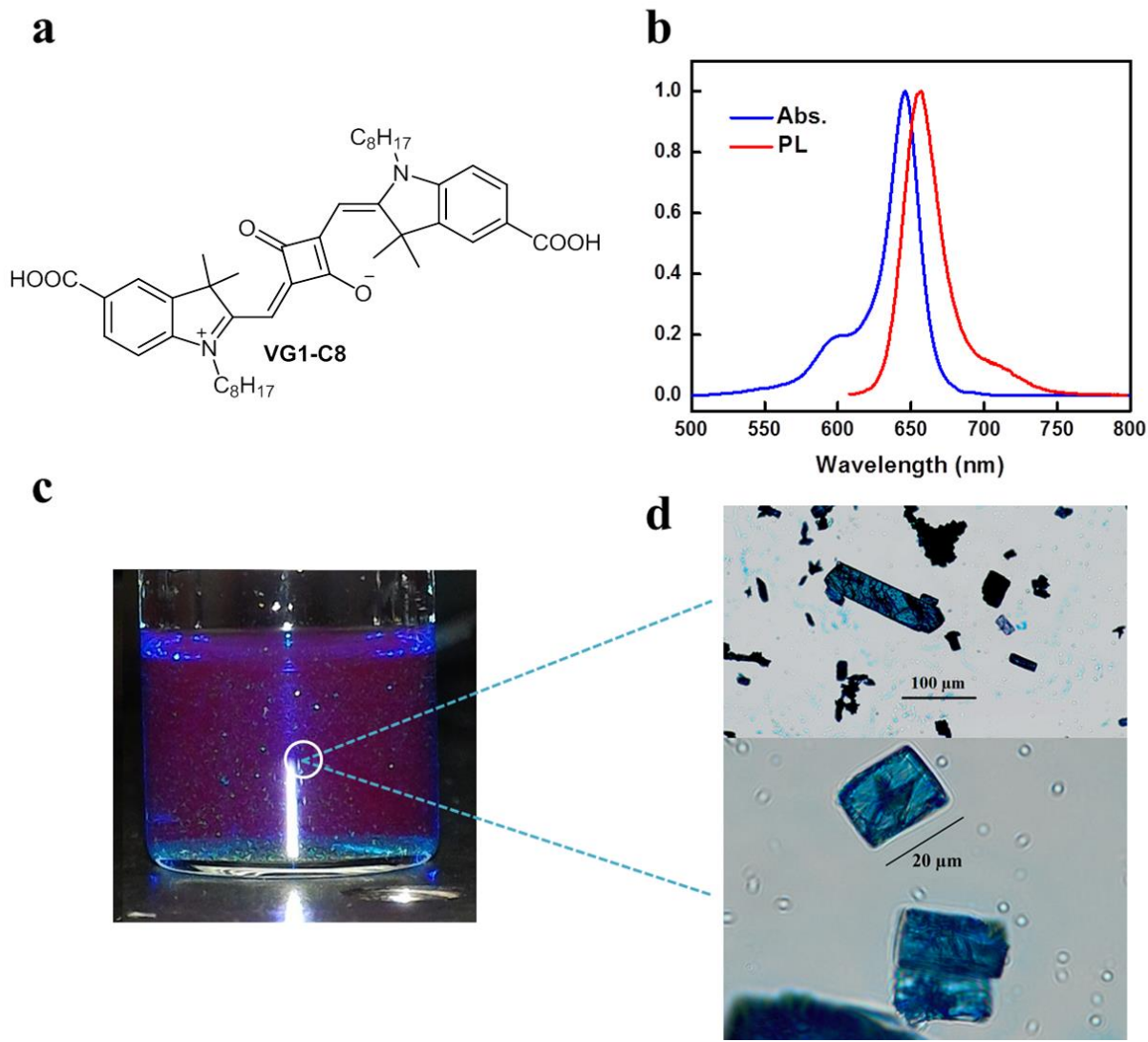

**Figure 1.** (a) Molecular structure of VG1–C8; (b) UV-Vis absorption and photoluminescence spectrum of VG1-C8 in ethanol solution (0.1 mg/mL); (c) spontaneous formation of shiny and aggregates in a 10 mg/mL ethanol solution of the dye; (c) optical microscope images of the aggregates.

In figure 2a-f we show the confocal transmission and PL maps of the aggregate. To acquire preliminary information about the transmission and PL spectral shape of the aggregate, we sampled the transmission at 500 nm, 690 nm and 800 nm, and the PL at 500 nm, 690 nm and 800 nm by using band-pass filters centred in those wavelengths (FWHM 10 nm). Starting from the transmission maps, we can observe that the position of the absorption peak is roughly centred at the same wavelengths of the solution absorption, even though the aggregate shows an extended transmission spectrum both at the lower (500 nm, fig. 2a) and higher (800 nm, fig. 2c) sides of the peak. We also note a decrease of the transmission in the area all-surrounding the aggregate that we attribute to the VG1-C8 film. Interestingly, the PL maps taken at 650 nm, 750 nm and 800 nm permit us to distinguish spatially three regions with strongly different spectral emission properties. In particular, by probing the PL at 650 nm we collect the fluorescence in the low-Stoke shift regime, which in our case comes mostly from the surrounding film. On the other hand, at 750 nm and 800 nm we sample mainly the emission from the border and the centre of the aggregate, respectively.

To summarise this section, the confocal transmission maps show an extended and broad absorption spectrum of the aggregate, which might be attributed to both intermolecular charge-transfer interactions in the solid state[23, 30] and/or to the presence of different domains with mixed J- and H- character. The PL maps, interestingly, show that emission from the aggregate is markedly red-shifted with the respect of the surrounding film, possibly indicating a higher order of intermolecular interactions than in film and in solution.[5]

**Confocal Transmission**

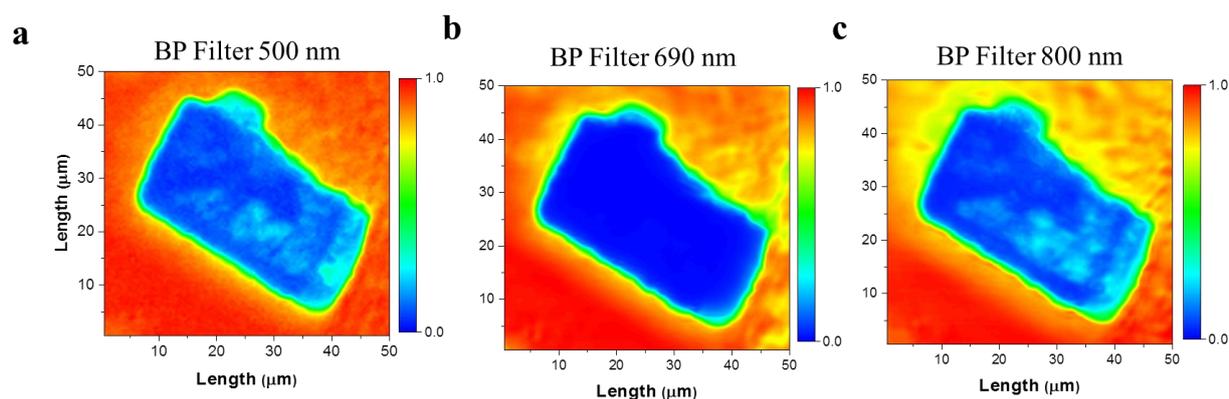

**Confocal PL**

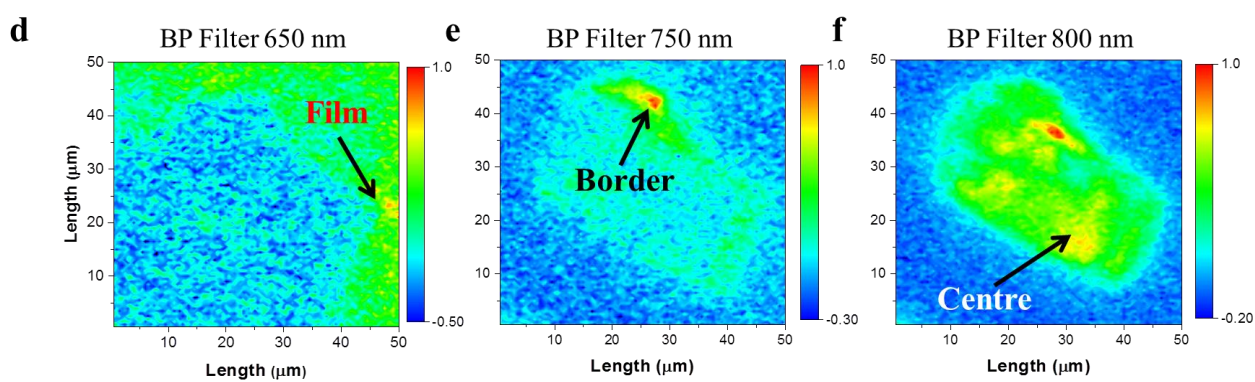

**Figure 2:** Normalised transmission (a-c) and PL (d-f) maps of the VG1 – aggregate (thickness ≈ 160 μm as measured by profilometry) and surrounding film (thickness ≈ 45 nm). We employed BP filters to resolve spatially and highlight the different spectral contributions.

Figure 3a-b shows the local normalised absorption (3a) and PL (3b) spectra taken in the three different regions (film, border and centre, see the arrows in figure 2d-f), as well as the absorption/PL spectra of the solution for comparison. We can note that the absorption of the film is broader and blue-shifted with respect of the solution spectrum, although still overlapping with the solution monomeric absorption (39 % of the total integrated area). We attribute this to a distribution of species namely: monomers, H-aggregates and possibly a low fraction of J- aggregates that could be below the sensitivity of our absorption measurements. If we pass to the border of the aggregate, we still see a blue-shift of the spectrum and an enhanced absorption at higher energy, which can be connected to the presence of a mixed region with a higher molecular disorder at the edge of the aggregate and to the generation of H-type species. However, we also observe that the overlap between the edge and monomeric absorption is considerably less than the previous case (27 %) and, in addition, there is the appearance of a new red-shifted band at 680 nm that can be a signature of an increased population of J-aggregates. Such findings may indicate that the fraction of monomeric VG1-C8 is substantially decreased at the expense of new generated H- and, to lower extent, J-aggregates. The scenario proposed above for the aggregate-edge lies somewhere in between the film and the aggregate core, as if we record the absorption spectrum at the centre we note that the monomeric peak lies almost in between the broad peak of the aggregate, which is concomitant with a decrease of the overlap with the monomeric band (24%). This can be linked to the coexistence of both J-type and H-type aggregates domains, or a dye arrangement that allows both J and H transitions (Davidov splitting).[18]

The PL spectra (fig. 3b), interestingly, seem to corroborate the aforementioned absorption results. In particular, the PL of the film is considerably overlapping with the solution emission (57 %), indicating the presence of a large fraction of monomeric species. However, we can note a slight red-shift of the emission (8 nm) and the appearance of a broad spectral feature lying at 732 nm (FWHM 113.9 nm) that can be attributed to a higher intermolecular order in the film and to the presence of J-aggregates, respectively. With regards to the aggregate-border, we observe the dramatic decrease of the monomeric peak together with the relative increase and narrowing of the component at 760 nm (FWHM 47 nm), which is consistent with the formation of emissive J-aggregates at the expenses of the monomeric form (24 % overlap). Remarkably, the aggregate core exhibits a mono-component, narrower and further red-shifted peak (780 nm, FWHM 45.3 nm) with a negligible overlap with the solution emission (5 %), indicating that the vast majority of the species is aggregated and, in particular, a significant fraction of these present a highly emissive J- configuration.

It is worth saying that the absolute amount of absorbed/emitted light within the aggregate will be dependent on the absorption cross sections and PL relative quantum yields of the different domains, as well as to thickness fluctuations within the aggregate (approx. 25 nm between the border and the aggregate centre) and self-absorption phenomena. Those effects in-fact hinder a quantitative characterisation of the absorption/emission properties of the aggregate, even though we consider that the profound spectral changes observed in the different domains, such as spectral shifts and variation of the relative magnitude of the various spectral components, are real and related to the different population of species (monomeric, H- and J- aggregates) in the investigated sample.

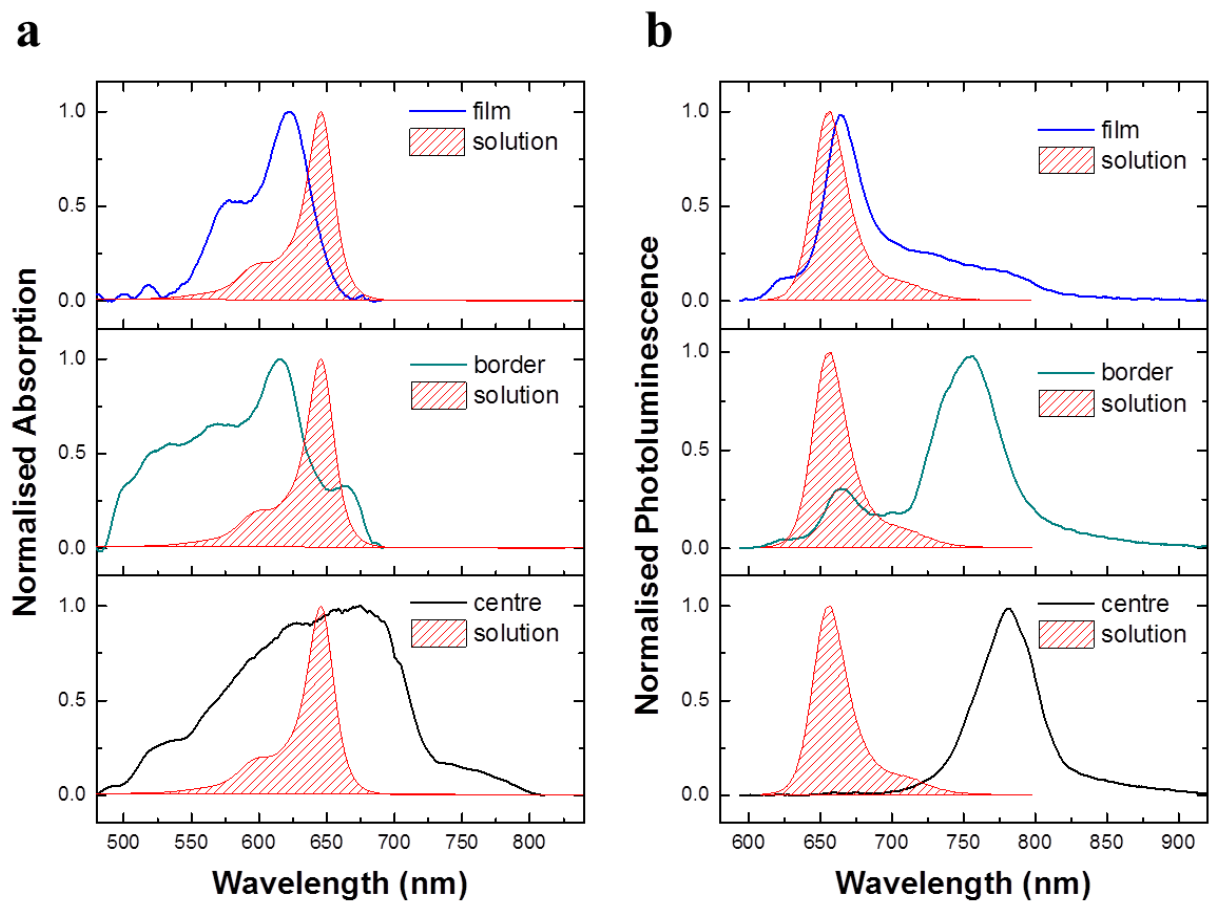

**Figure 3:** Absorption (a) and PL (b) spectra taken at the three different regions namely, film, aggregated-border and centre. The absorption and PL spectra of a diluted solution in ethanol of VG1 –C8 (0.14 mM ) is also reported for comparison.

## Conclusions

In summary, we have investigated the absorption/emission properties of a single squaraine aggregate by means of confocal microscopy, UV-Vis and PL spectroscopy. By resolving spatially and spectrally the transmission and emission of the aggregate and taking the local absorption/emission spectra, we observe that the aggregate core shows a broad absorption band centred at 670 nm likely originating from the presence of Davidov aggregates with allowed J- and H- configuration. Interestingly, such a region exhibits a relatively narrow emission in the NIR part of the spectrum with little overlap with the broadband absorption spectrum (Stokes shift 110 nm). The broad absorption spectrum, the strong and relatively narrow NIR emission alongside the small spectral overlap between absorption and PL could make these aggregates, if conveniently processed and manipulated, greatly appealing for the fabrication of efficient photovoltaic and light-emitting diodes, as well as NIR probes in biomedical applications.

## Acknowledgements

This work is supported by the H2020 ETN SYNCHRONICS under grant agreement 643238.